\documentclass{ar-1col-S2O}
\usepackage[numbers]{natbib}
\usepackage{url}
\usepackage{amsmath,amssymb, esint}

\usepackage{array}
\usepackage{tabularx}
\newcolumntype{x}[1]{>{\centering\arraybackslash}m{#1}}
\newcolumntype{Y}{>{\centering\arraybackslash}X}

\usepackage{hyperref}
\usepackage{url}

\usepackage{comment}

\allowdisplaybreaks

\jname{Annu. Rev. Condens. Matter Phys.}
\jvol{AA}
\jyear{YYYY}
\doi{10.1146/((please add article doi))}

\begin{document}

\markboth{Classen and Betouras}{High-order Van Hove singularities and flat bands}

\title{High-order Van Hove singularities and their connection to flat bands}

\author{Laura Classen$^{1,2}$ and Joseph J. Betouras$^3$
\affil{$^1$Max-Planck-Institute for Solid State Research, Stuttgart 70569, Germany}
\affil{$^2$Department of Physics, Technical University of Munich, Garching 85749, Germany}
\affil{$^3$Department of Physics and Centre for the Science of Materials, Loughborough University, Loughborough LE11 3TU, UK}}

\begin{abstract}
The flattening of single-particle band structures plays an important role in the quest for novel quantum states of matter due to the crucial role of interactions. 
Recent advances in theory and experiment
made it possible to construct and tune systems with nearly flat bands, ranging from graphene multilayers and moir\'e materials to kagom\'e metals and ruthenates. 
While theoretical models predict exactly flat bands under certain ideal conditions, evidence was provided that 
these systems host high-order Van Hove points, 
i.e., points of high \emph{local} band flatness and power-law divergence in energy of the density of states. In this review, we examine recent developments in engineering and realising such weakly dispersive bands. We focus on high-order Van Hove singularities and explore their connection to exactly flat bands. We provide classification schemes 
and discuss interaction effects.   We also review experimental evidence for high-order Van Hove singularities and point out future research directions. 
\end{abstract}

\maketitle

\tableofcontents

\section{INTRODUCTION: definition and importance of high-order Van Hove singularities}

Different phases of matter driven by non-trivial topology and geometry of electronic band structure 
are of enormous interest for understanding and controlling correlated quantum systems. 
While various effects of topology have been well studied and documented, new avenues continue to emerge in the investigation of band structure geometries and the underlying symmetry properties of a system. Pioneering early work by Lifshitz and Van Hove \cite{Lifshitz_1960,VanHove} laid the foundation for a rich path of uncovering the exotic effects of Fermi surface geometry and 
the consequences 
of Fermi surface topological transitions where the Fermi surface geometry undergoes a sudden change when some parameters in the system are changed. Lifshitz initially studied two particular forms of these topological changes (pocket appearing/disappearing or neck formation/collapse)~\cite{Lifshitz_1960, Abrikosov_1988}. 
In general, such transitions happen when the Fermi surface hosts critical points of the dispersion.  

A critical point of the energy dispersion appears when the gradient of the dispersion $\varepsilon_n({\bf k})$ of the $n^{th}$ band at some point ${\bf k}_0$ 
vanishes $\nabla \varepsilon_n ({\bf k_0})=0$. 
The critical point can be a maximum, minimum, or a saddle. 
It is called non-degenerate if the determinant of the Hessian does not vanish simultaneously $\det[\partial^2\varepsilon_n({\bf k})/ \partial k_\mu \partial k_\nu]\neq 0 $.  
In this case, 
the dispersion around the 
critical point can be described with a Taylor expansion to quadratic order, taking a canonical form $\pm k_x^2 \pm k_y^2\pm \ldots \pm k_d^2$ in $d$ dimensions. 
Van Hove showed that the dispersion of a crystalline system in the relevant case of two dimensions (2D) must always possess a saddle point, which is 
accompanied by a logarithmic divergence in the density of states (DOS)~\cite{VanHove} at the corresponding energy. 
Notably, there 
is a class of critical points around which the dispersion needs to be Taylor expanded beyond quadratic order 
because they are degenerate $ \det[\partial^2 \varepsilon_n({\bf k})/ \partial k_\mu \partial k_\nu] = 0 $. 
They are known as high-order critical points, and the corresponding 
DOS displays a high-order Van Hove singularity (HOVHS). 

At a high-order critical point, the Fermi surface becomes singular as happens at an ordinary Van Hove singularity (VHS). But HOVHS are 
accompanied by power-law diverging DOS, especially in 2D~\cite{Chandrasekaran_PRR_2020, LiangFu}.
HOVHS have a strong impact on a system's properties when close to the Fermi energy (E$_F$). 
While a direct signature of the HOVHS
can, e.g., be observed 
in the tunnelling conductivity~\cite{Yuan2019, magic_VanHove2}, 
the singular DOS also affects 
thermodynamic and transport quantities, as well as
all kinds of susceptibilities.
Furthermore, electronic correlations 
are strongly enhanced by the large DOS in the vicinity of 
a HOVHS 
so that interaction-induced many-body phases are expected to emerge.

Traditional Lifshitz  transitions and associated in 2D logarithmically-divergent VHS were reported in a variety of materials including
cuprates, iron-based superconductors, cobaltates, $\text{Sr}_{2} \text{RuO}_{4}$ and
heavy fermions~\citep{Aoki, Barber, Bernhabib, Khan, Coldea, Okamoto, Sherkunov-Chubukov-Betouras, Slizovskiy-Chubukov-Betouras, Yelland}.
With the experimental capabilities being improved continuously, there is a 
recent surge of interest in HOVHS~\cite{HOvHS1,HOvHS2,HOvHS3,HOvHS4,HOvHS5}. 
Signatures of HOVHS were found in many materials which show correlated electron behavior.  
They include ruthenates, such as $\text{Sr}_{3} \text{Ru}_{2} \text{O}_{7}$
where a
HOVHS was shown to exist in the presence
of an external magnetic field~\cite{Efremov_PRL_2019, Chandrasekaran_PRR_2020}, or Sr$_2$RuO$_4$, where a HOVHS can occur on the surface \cite{chandrasekaran2023engineering}.
Evidence for a different HOVHS was reported in highly overdoped graphene~\cite{Rosenzweig_2020} and kagom\'e metals \cite{Kang2022, Hu2022}. 
In twisted bilayer graphene \cite{Cao2018, Cao2018SC}, they were discussed in the context of the occurrence of almost flat bands around a so-called magic twist angle \cite{Yuan2019}. Furthermore, they
were suggested to be relevant for the recently observed phases of Bernal bilayer graphene~\cite{seiler2022,Zhou_2021}. 

In this review, we aim at (i) presenting the recent developments of the area in a systematic way and (ii) discussing methods to engineer and realise these singularities, making the connection to flat bands as one of these methods. For (ii) we present a non-exhaustive review of the recent work on flat bands. 
In section 2, we discuss ways to make HOVHS accessible and show their connection to flat bands. Section 3 is devoted to classification and response of these HOVHS, giving emphasis on the two-dimensional (2D) cases and discussing in passing one (1D) and three dimensions (3D). In section 4 we describe interaction effects, which are pronounced, and section 5 provides evidence for their experimental realisation in quantum materials, discussing three main categories of prime importance: strontium ruthenates, kagom\'e metals, and 
graphene and moir\'e materials. We finish with an overview and suggestions for directions to uncover more of the fascinating physics.

\section{ACCESSIBILITY OF HOVHS AND FLAT BANDS}
\subsection{Tuning from ordinary to high-order}
\label{sec:tuningtoHOVHS}

In a single-particle picture, HOVHS typically require delicate tuning of parameters in the system to obtain (effected through strain, pressure, twisting angle, magnetic field, gate and bias voltage, etc). This has lately been made possible with the advances in experimental techniques. Although infinitely many such distinct singularities exist, catastrophe theory~\cite{castrigiano} guarantees that we are typically likely to obtain only a finite subset of singularities in real systems. The maximal number of tuning parameters needed is restricted by the type of the singularity and the symmetry of the system (see Sec.~\ref{sec:classes}). Note, however, that self-energy effects can favor the formation of a HOVHS beyond the single-particle picture (see Sec.~\ref{sec:selfenergy}).

\begin{figure}[t]
	\center{\includegraphics[width=.99\linewidth]{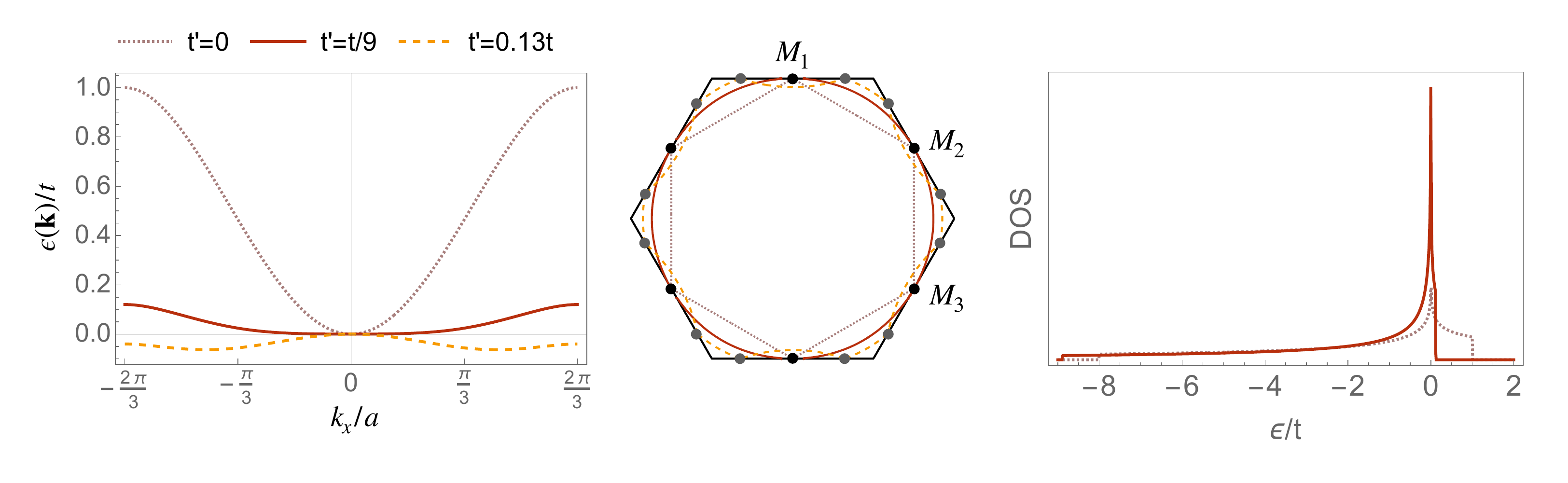}} 
	\caption{Triangular-lattice energy dispersion at $(k_x,2\pi/\sqrt{3})/a$ going through 
 Van Hove point 
 $M_1$ at $k_x=0$ for three values of the next-nearest-neighbor hopping $t'/t$ and corresponding Fermi surface at Van Hove filling. A HOVHS occurs for $t'=t/9$ (red solid line). For small $t'\leq t/9$ (rose, dotted), there are 
 three Van Hove points at the $M_i$ points of the Brillouin zone (black dots). Each splits into two (gray dots) located along $K'-M-K$ for $t'>t/9$ (dashed, orange). The density of states (DOS) shows a stronger power-law singularity in the case of a HOVHS.}
	\label{fig:triag}
\end{figure}

Mathematically, HOVHS are unstable to certain classes of perturbations that cause the high-order critical points to break into a number of ordinary critical points determined by their multiplicity \cite{LiangFu}. Vice versa, a strategy to tune towards a HOVHS is starting from an ordinary saddle point or a set of ordinary critical points and varying some tuning parameter(s) in such a way that the dispersion relation becomes locally flat, i.e. not only its linear but also its quadratic part vanishes along at least one direction. To illustrate this, let us consider the triangular lattice with nearest- and next-nearest-neighbor hopping as a minimal toy model  
\begin{align}
    H_0=-t \sum_{\langle i,j\rangle} (c_i^\dagger c_j + \mathrm{h.c.})- t'\sum_{\langle \langle i,j \rangle\rangle} (c_i^\dagger c_j + \mathrm{h.c.}) - \mu\,.
\end{align}
The chemical potential $\mu$ is used to control the Fermi level and we assume that the ratio $t'/t$ can be varied by some external parameter. 
For example, it was proposed that the triangular-lattice Hubbard model can be simulated via twisted transition metal dichalcogenides \cite{Fengsheng2018}. The dispersion relation of $H_0$ is given by $\epsilon({\bf k})=-2t \sum_{n=1}^3\cos ({\bf k}\cdot {\bf a}_n)-2t' \sum_{n=1}^3\cos ({\bf k}\cdot {\bf b}_n)$, where ${\bf a}_n$ and ${\bf b}_n$ are three independent nearest- and next-nearest-neighbor vectors. For concreteness, we use ${\bf a}_1=(1,0)a,{\bf a}_2=(1,\sqrt{3})a/2,{\bf a}_3=(1,-\sqrt{3})a/2$ and ${\bf b}_1=(0,\sqrt{3})a,{\bf b}_2=(3,\sqrt{3})a/2,{\bf b}_3=(3,-\sqrt{3})a/2$ with lattice constant $a$. For small values of $t'<t'_c$, there are 
three nonequivalent saddle points at the M points of the Brillouin zone (see Fig.~\ref{fig:triag}). Moving the Fermi level to the saddle points $\mu=2(t+t')$ and expanding around them, e.g. around ${\bf M}_1=(0,2\pi/\sqrt{3})$, we obtain up to quartic order 
\begin{align}
    \epsilon({\bf M}_1+{\bf k})&=\frac{1}{2}(t-9t')k_x^2-\frac{3}{2}(t-t')k_y^2 \nonumber \\ 
    &+\frac{1}{96}(-7t+81t')k_x^4 + \frac{3}{32}(t-7t')k_y^4 + \frac{3}{16}(t+9t')k_x^2k_y^2 + \mathcal O (k^6)\,.
\end{align}
It is easy to see that, upon tuning $t'/t$, the first critical value where a leading-order term vanishes is $t'_c=t/9$. At this point the leading order terms of the expansion are 
\begin{align}\label{eq:HOdispM1}
    \epsilon_c({\bf M}_1+{\bf k})&= -\frac{4}{3}t k_y^2 +\frac{1}{48}t k_x^4 + \frac{3}{8}t k_x^2 k_y^2\,,
\end{align}
which represents a HOVHS (cusp) that contains the canonical part $\sim k_x^4-k_y^2$ (after rescaling) and the irrelevant perturbation $\sim k_x^2 k_y^2$. 
The DOS at this HOVHS diverges as $|\delta E|^{-1/4}$, which does not change when higher-order terms are included in the Taylor expansion. Increasing $t'$ beyond $t_c$, each saddle point splits into one maximum at the M points and two saddle points moving towards K and K', leaving a total of 
six VHS in the Brillouin zone \footnote{The next critical value for $t'$ is $t'_{c,2}=t/6$, where three saddle points merge with a maximum at the K (K') point and form another type of HOVHS with canonical dispersion $3k_x k_y^2-k_x^3$. This can be seen by expanding around the K (K') points.}. Thus, reaching the HOVHS requires fine-tuning of parameters ($t'$ in our example). Nevertheless, the behavior of the system will be strongly influenced by the HOVHS in a finite range around the special parameter set. 

\subsection{Connection to flat bands}
\label{sec:flatbands}

Closely related in spirit to the interest in HOVHS 
is the study of engineering flat bands in an energy spectrum. In both cases, fine-tuning potentially facilitated by symmetries is supposed to achieve (local or global) band flattening which promises a rich phenomenology due to strong interaction effects. 

An exact flat band signifies the presence of localised states in real space, which can be rooted in different mechanisms. Two distinct physical ways leading to the formation of dispersionless bands in lattices are (i) localisation of wave functions in real space resulting in zero hopping elements and isolated atomic orbits and (ii) quantum interference effects where destructive interference leads to flat bands. In continuum models, flat bands corresponding to Landau levels can also arise. 
Here we will not consider case (i) without hopping, because we are interested in a connection to HOVHS. 
In case (ii), flat-band constructions from wavefunction interference are usually constrained by requirements on symmetry or lattice geometry so that perturbations give rise to a finite, albeit small dispersion. As a remnant of the formerly flat band, a HOVHS can emerge. While the circumstances when this happens are still subject of current research, we 
demonstrate this phenomenon for a general class of flat bands  
below. 
A broader motivation for investigating the connection of HOVHS and flat bands comes from the implied existence of a limit where all orders of a local Taylor expansion around a critical point must vanish. Then, a perturbed system where the flat-band conditions are violated still provides a convenient starting point for fine-tuning to achieve the less restrictive requirement of local band flattening where just two terms of the expansion must vanish.

\subsubsection{Flat bands and HOVHS in lattice systems}

In an early example of a flat band in a dice lattice, it was proposed that the existence of localized states may depend only on the local topology of the hopping matrix elements \cite{Sutherland_1986}. 
More generally, flat bands due to destructive interference 
are found in lattices with geometric frustration in the hopping elements, such as the kagom\'e lattice \cite{Bergman_Wu_Balents_PRB_2008}. This can be related to theorems of graph theory. For tight binding models with nearest neighbor hopping on lattices which are the  
line or split graph of another lattice,
they guarantee that every such lattice has in its energy spectrum at least one flat band \cite{Kollar_Houck_etal_2020,Ma_split2020,Lieb1989}. 

A lattice is constructed via a line graph by replacing each bond of a lattice by a site and connecting sites on bonds that shared a site in the original lattice. For a split-graph construction, an additional site is added on each bond of a lattice. 
For example the line graph of a hexagonal lattice is the kagom\'e lattice and the spectrum of a tight binding model with only nearest neighbor hopping 
of the kagom\'e lattice possesses a flat band; similarly for the checkerboard lattice which is the line graph of the square lattice. These statements hold also in the case of non-Euclidean lattices \cite{Kollar_Houck_etal_2020}. An example for a split graph is the Lieb lattice, constructed from the square lattice. 

Furthermore, it was 
shown in different lattice models with more than one band, such as the Haldane model on a honeycomb lattice or a checkerboard lattice with time-reversal symmetry breaking, that further-neighbor hopping can stabilize flat bands with non-trivial Chern numbers\cite{Neupert_PRL_2011, Sun_Gu_Katsura_DasSarma_PRL_2011}. 
Different ways to construct flat bands have been extensively reported e.g. \cite{Ramachandran_PRB_2017, Bi_Yuan_Fu_PRR_2019, Xu_Bu_PRA_2020}.
Generalising some of these ideas, a technique for constructing perfectly flat bands in bipartite crystalline lattices with unequal numbers of orbitals was recently developed, and used for a complete topological classification of flat bands \cite{Calugaru2022}.

An interesting question in this context is under which topology and symmetry requirements 
an isolated (almost) flat band can be created in a lattice system.  
As a next step, we are then
interested in a small breaking of these requirements 
to analyse if this can, in principle, produce a HOVHS. 

Flat bands are generally made of compact localised states. 
These state are connected to local symmetry transformations \cite{Roentgen_PRB_2018}. However, a flat band can also  
contain delocalised states due to touching with a dispersive band. In particular, localisation can be prohibited because of non-trivial topology. 
If the compact localised states can form a complete set spanning a flat band or need to be complemented by extended states depends on the behavior of its Bloch wave function in momentum space. It is categorised as singular if the Bloch wave function possesses immovable discontinuities generated by band-crossing with other bands, and as non-singular if there are no such discontinuities \cite{Rhim_Yang_PRB_2019}. A non-singular flat band can be completely isolated from other bands while preserving the perfect flatness. In contrast, a singular flat band also contains extended states and cannot form an isolated flat band. It is flat as long as there is a band touching with another band. If perturbations lift the degeneracy of the band touching, the flat band must become dispersive. It can acquire a Chern number in this process, providing a mechanism for the creation of an isolated, nearly flat Chern band. For example, nearly flat bands with nonzero Chern numbers are produced by opening a gap at a non-trivial quadratic band touching on a square or checkerboard lattice \cite{Sun_Gu_Katsura_DasSarma_PRL_2011,Sun_Yao_Fradkin_Kivelson_PRL_2009}.  
Instead of opening a gap, a perturbation can also shift the flat band in the other direction into the dispersive band. 

The class of singular flat bands can be described by a 
general continuum model 
in two dimensions \cite{Rhim_Yang_PRB_2019}. We will use this to demonstrate how a HOVHS arises from perturbing the flat band. To lowest order in momentum, a band touching with a singular flat band is necessarily quadratic (in the case of a linear touching, the flat band is always non-singular) \cite{Rhim_Yang_PRB_2019}. The most general form of the Hamiltonian for this case is:
\begin{align}\label{eq:genericsingular}
H_k= h_0({\bf k})\sigma_0 + \sum_i h_i({\bf k})\sigma_i\, 
\end{align}
with Pauli matrices $\sigma_{x,y,z}$, identity matrix $\sigma_0$, and where $h_\mu$ are parameterised as 
$h_x= t_6 k_y^2$, $h_y=t_4 k_x k_y + t_5 k_y^2$, $h_z= t_1 k_x^2+t_2 k_x k_y + t_3 k_y^2$, and $h_0=b_1 k_x^2 +b_2 k_x k_y +b_3 k_y^2$.
Only four of the parameters $t_i,b_i$ are independent because of the flat-band condition $\det H_k=0$ and without loss of generality we consider $t_1>0$. Adding a small perturbation $m \sigma_{i}$, $i=x,y,z$ to Eq.~\eqref{eq:genericsingular} can either produce a gap or split the quadratic band touching into two Dirac points. 
In the latter case, a HOVHS is created at the origin between the Dirac points \footnote{In the first case, one needs to go beyond the continuum model to study Van Hove singularities.}. This can be seen most easily by adding a mass $m \sigma_z$ with $m>0$ as a perturbation. The resulting eigenenergies are $\epsilon^\pm_{\bf k}=h_0({\bf k})\pm\sqrt{h_x^2({\bf k})+h_y^2({\bf k})+(h_z({\bf k})+m)^2}$, which reduces to $\epsilon(k_x,0)=b_1 k_x^2\pm|m+t_1 k_x^2|$ in $k_x$ direction. Using 
one of the flat band conditions 
$b_1^2=t_1^2$, we obtain that one of the bands remains dispersionless in $k_x$ direction, e.g., if $b_1=t_1$ we get $\epsilon^-(k_x,0)=-m$. At the same time, the band acquires a dispersion in $k_y$ direction. Thus, while the perturbation destroys the complete flatness of the band, it remains robust along one direction. The result is a special HOVHS with complete flatness in this direction.

\subsubsection{Flat bands and HOVHS in continuum Dirac models}
\label{flatDirac}
A further, intensely discussed route towards band flattening applies to systems harboring Dirac fermions where the Dirac velocity is reduced to zero. In perhaps the most prominent example of twisted bilayer graphene, this is achieved through tuning the twist angle between layers \cite{BMmodel}. Similarly, the Dirac velocity can be manipulated, e.g., in twisted few-layer graphene \cite{Khalaf2019,Carr2020,Zhu2020}, twisted nodal superconductors \cite{Volkov2023,Can2021,Song2022}, and twisted surfaces of topological insulators \cite{Cano2021,Wang2021,Dunback2022}. 
There are many extensions to other systems and other tuning parameters based on the general idea of band manipulation via the introduction of a periodic potential modulation (coming from the moir\'e pattern in the case of a twist). 
For example, this includes systems without Dirac fermions, such as twisted transition metal dichalcogenides \cite{Fengsheng2018,Fengcheng2019,Pan2020,Zang2021} or twisted quadratic band touchings \cite{Li2022}, as well as systems without a twist, such as graphene with impurity arrangements \cite{Ahmadkhani_PRB_2023} or periodic substrate potential \cite{Lu_AdvMat_2022}. 
This lead to the general class of moir\'e materials \cite{Balents2020,moiremarvels,Kennes2021}, where it is likely to find HOVHS due to their high tunability. 

Focusing on the case of Dirac fermions, the symmetry requirements and number of tuning parameters necessary for the vanishing of the Dirac velocity were worked out in Ref.~\cite{Sheffer_PRX_2023}. In many cases of interest, the symmetry of a system (e.g. a rotation symmetry) fixes the position of Dirac points or prohibits a quadratic order of the band dispersion. Then the lowest-order dispersion must become cubic when the Dirac velocity vanishes, which guarantees a HOVHS with a power-law divergence in the DOS at least of order $\delta E^{-1/3}$. Furthermore, in extreme limits the vanishing of the Dirac velocity implies the emergence of an exactly flat band. 
Such limits can arise in Dirac Hamiltonians with an external SU(2) gauge field. 
The states forming the corresponding flat bands resemble Landau level wave functions and have a non-zero Chern number~\cite{Sheffer_PRX_2023, Tarnopolsky2019}. 
In the case of TBG, it was shown that an approximate particle-hole symmetry needs to be exact for the
vanishing of the Dirac velocity, and an additional chiral symmetry needs to be present for the complete flattening. 
Yet, an exceptional local flattening of the bands in the form of HOVHS was demonstrated to occur without these symmetries \cite{Yuan2019}. The rationale is that, when the moir\'e bands have a finite bandwidth, they must  possess a VHS. The number 
of associated saddle points changes from three to six per valley as function of the angle. Upon approaching the magic angle,  
there is a critical value, when each Van Hove point splits into two (plus one extremum), similarly to the example given in Sec.~\ref{sec:tuningtoHOVHS} above.
At the critical angle,  the VHS is of high order with a quartic momentum dependence $k_i^4$ 
in one direction. The associated divergent density of states with scaling $\delta E^{-1/4}$ 
fits the measured tunneling conductance well \cite{Yuan2019} (see Sec.~\ref{sec:graphenemat}).

\section{CLASSIFICATION OF HOVHS}
\label{sec:classes}
\subsection{Characterising indexes}
Critical points of the dispersion $\epsilon(\bf k)$, and in particular high-order critical points, strongly affect physical properties because they lead to a non-analytic DOS. This can be systematically classified based on on a Taylor expansion around the critical point \cite{Chandrasekaran_PRR_2020, LiangFu}. 

A mathematical framework for the classification comes from catastrophe theory, which analyzes degenerate points ($\nabla \epsilon(\bf k)=0$, $ \det[\partial^2 \epsilon({\bf k})/ \partial k_\mu \partial k_\nu] = 0 $) of real-valued, differentiable functions.  
Different types of HOVHS can be represented by a typical form of their Taylor expansion plus a small number of perturbations,   
known as the \emph{universal unfolding} in the language of catastrophy theory. 
HOVHS of the same class are equivalent up to smooth coordinate transformations. 
The representatives, in turn, can be classified by four integer indexes corresponding to corank, codimension, determinacy, and winding \cite{Chandrasekaran_PRR_2020}. 
The \emph{corank} is the number of zero eigenvalues of the Hessian matrix - the matrix of second derivatives of the function $\{\partial^2 \epsilon({\bf k})/ \partial k_\mu \partial k_\nu\}_{\mu,\nu}$ - at the critical point. 
The \emph{codimension} corresponds to the 
finite number of effective control parameters modulating the polynomial perturbations in the universal unfolding. 
These polynomial perturbations 
are distinguished by their behavior under smooth coordinate transformations which defines 
directions in polynomial space which are not tangent to the orbit of the high-order singularity.  
In a physical system, some of the control parameters can be required to be zero due to symmetry, effectively reducing the number needed to reach the critical point. 
Finally, 
the \emph{determinacy} describes the truncation order of the Taylor expansion characteristic for a HOVHS. 
The above three indices can classify singularities in any dimension but they cannot resolve all degeneracies. 
Focusing on functions of two variables 
for the case of 2D electron dispersions, 
degeneracies 
are resolved 
through the last index, 
the \emph{winding} number $w$, which counts the sign changes of the function in a small loop around the 
critical point.

The different classes that describe the asymptotic behavior close to a critical point 
are also related to a topological index and scaling exponents \cite{LiangFu}. 
The unperturbed dispersion of a HOVHS scales according to  $\epsilon(\lambda^{a_i}k_i)=\lambda \epsilon(k_i)$, ($i=1,\ldots,D$). The scaling exponents $a_i$ further determine the multiplicity (Milnor number) of the critical point and the behavior of the DOS. A critical point of multiplicity $\mu$ can split at most into $\mu$ critical points, which also means that there are at most $\mu-1$ independent relevant perturbations. 
The topological index employed in this case is 
\begin{equation}
I= \begin{cases}
\frac{1}{2} [\mathrm{sgn}\,\upsilon(+\delta)-\mathrm{sgn}\,\upsilon(-\delta)], &in\;\;\; 1D\\
\frac{1}{2\pi} \mathrm{Im}\oint_{\mathcal{C}} \frac{d\upsilon_x+id\upsilon_y}{\upsilon_x+i\upsilon_y}, &in\;\;\; 2D\\
\frac{1}{4\pi} \oiint_{\mathcal{S}} \hat{\upsilon} \cdot \frac{\partial \hat{\upsilon}}{\partial k_{\parallel}} \times \frac{\partial \hat{\upsilon}}{\partial k_{\perp}} dk_{\parallel} dk_{\perp}, &in\;\;\; 3D
\end{cases}
\;,
\end{equation}
where the velocity field $\upsilon=\frac{\partial \epsilon}{\partial k}$ in 
1D and ${\bf \upsilon} = \nabla_{\bf k}\epsilon$ in 2D and 3D. The index $I$ is a winding number (different from $w$) around the critical point at $k=0$. 
The interval $[-\delta, \delta]$ in 1D, the contour $\mathcal{C}$ taken counterclockwise in 2D, and the surface $\mathcal{S}$ in 3D all enclose the only critical point at $k=0$. 
With this definition, the Poincar\'e-Hopf theorem ensures that $\sum_i I_i =0$, where $i$ enumerates all critical points in the Brillouin zone. 
The reason is that the total topological index of all critical points in 
momentum space reflects the topology of the toric
Brillouin zone, equal to its Euler characteristic of zero.

\subsection{Classification schemes (especially in 2D)}
Using catastrophe theory, 
all possible HOVHS that can occur in the electronic dispersion of
electrons in 2D were classified in Ref.~\cite{Chandrasekaran_PRR_2020}. 
Catastrophe theory guarantees that typically only
high-order singularities with codimension (or cod) $\leqslant 7$ occur if there are at most seven control parameters. 
There are 17 singularities with cod $\leqslant 7$.  
As previously mentioned, crystal symmetries are likely to reduce the effective number of necessary control parameters drastically. This is why another 
atypical singularity called $X_9$ was also provided, which has codimension $8$. 
It is the singularity with the lowest codimension that respects four-fold rotational symmetry. Thus, in a tetragonal system, 
many of the constraints are automatically satisfied.

One can further combine the result of catastrophe theory with symmetry considerations.  
In 2D, there are seventeen wallpaper groups which serve as symmetry groups of crystals. 
In addition, there is inversion symmetry in k-space for time-reversal invariant systems without spin-orbit coupling. 
By combining point group elements with the reciprocal lattice translations, high-symmetry points of the Brillouin zone were 
identified and the allowed type of HOVHS at each of these points was listed \cite{Chandrasekaran_PRR_2020}. 
Similarly,
the different types of HOVHS were listed in Ref.~\cite{LiangFu} specifying their canonical dispersion and relevant perturbations  depending on topology, scaling, and symmetry around the critical point. 

Each of the classes also implies a universal behavior of the DOS. 
As we approach a critical
point in momentum space, the 
DOS 
also approaches 
a singularity in energy, described by a scaling law
specific to the critical point  
\begin{equation}
\rho(\epsilon) = \begin{cases}
D_+ |\epsilon|^{-\gamma}, &\epsilon > 0\\
D_- |\epsilon|^{-\gamma}, &\epsilon < 0
\end{cases}
\;. 
\end{equation}
The exponent $\gamma=\sum_i a_i-1$ can be inferred from the scaling exponents of the dispersion. 
In addition, 
it was 
shown \cite{Chandrasekaran_PRR_2020, LiangFu} that 
the ratio of the coefficients are universal, 
i.e., preserved under smooth coordinate transformations. 
The DOS of a HOVHS can be particle-hole asymmetric $D_+/D_-\neq 1$.  
In contrast, at an ordinary VHS, the DOS diverges in a logarithmic $\gamma=0$, particle-hole symmetric $D_+/D_-=1$ fashion. 
We reproduce common types of HOVHS and their properties in Tab.~\ref{table1}. 

Even given a classification, 
the detection of HOVHS in multi-band systems 
can be a formidable task in practice. 
However, 
a quantitative method to detect, diagnose, and engineer HOVHS in multiband systems 
for any 
number of bands or hopping terms has been developed \cite{Chandrasekaran_Betouras_2023}. 
Given a $\bf k$-space Hamiltonian,  
the Taylor expansion of the dispersion of any band at arbitrary points in momentum space 
can be computed, 
using a generalized extension of the Feynman–Hellmann theorem.  
As a result, the presence and tunability of a HOVHS for a system can be analysed, and a low-energy theory in the vicinity of the point of interest in the Brillouin zone can be constructed.

\begin{table}
\caption{List of example catastrophes that classify HOVHS relevant to 2D lattice systems with a low number of tuning parameters. They are uniquely indexed by corank (cr), codimension (cd), determinacy (d) and winding (w). The universal unfolding represents the canonical dispersion expanded around the wave vector where it vanishes. Relevant perturbations are $\propto t_i$. Each class implies a singular DOS, which diverges in a power-law fashion with a universal exponent and a universal ratio $D_+/D_-$ for its coefficients.}
\label{table1}
\footnotesize
\begin{center}
\begin{tabularx}{\textwidth}{|@{ }x{1.7cm}@{ }|@{ }x{1.7cm}@{ }|@{ }x{1cm}@{ }|@{ }x{1.3cm}@{ }|Y|@{}x{1.9cm}@{ }|}
\hline
\begin{tabular}{@{}c@{}}
Catastrophe \\
(ADE index)
\end{tabular}
&
(cr, cd, d, w) 
&
DOS
&
$D_+/D_-$
&
\begin{tabular}{@{}c@{}}
Universal \\
unfolding
\end{tabular}
&
\begin{tabular}{@{}c@{}}
Energy \\ contours 
\end{tabular}
\\
\hline
\begin{tabular}{@{}c@{}}
Fold\\
($A_2$)
\end{tabular}
&
$(1,1,3,2)$
&
$|\epsilon|^{-1/6}$
&
$\frac{1}{\sqrt{3}}$
&
\begin{tabular}{@{}c@{}}
$k_x^3-k_y^2 + t_1 k_x$
\end{tabular}
&
\begin{tabular}{@{}c@{}}
\includegraphics[scale=0.5]{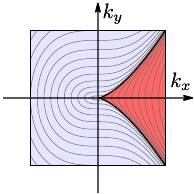}
\end{tabular}\\
\begin{tabular}{@{}c@{}}
Cusp\\
($A_3$)
\end{tabular}
&
$(1,2,4,4)$
&
$|\epsilon|^{-1/4}$
&
$\frac{1}{\sqrt{2}}$
&
\begin{tabular}{@{}c@{}}
$k_x^4-k_y^2 + t_2 k_x^2 + t_1 k_x$
\end{tabular}
&
\begin{tabular}{@{}c@{}}
\includegraphics[scale=0.5]{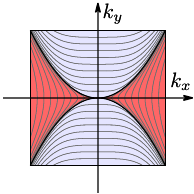}
\end{tabular}\\
\begin{tabular}{@{}c@{}}
Elliptic\\
umbilic\\
($D_4^-$)
\end{tabular}
&
$(2,3,3,6)$
&
$|\epsilon|^{-1/3}$
&
$1$
&
\begin{tabular}{@{}c@{}}
$3k_x^2 k_y-k_y^3$ \\
$+t_3 (k_x^2 + k_y^2)+t_2 k_y +t_1 k_x$
\end{tabular}
&
\begin{tabular}{@{}c@{}}
\includegraphics[scale=.5]{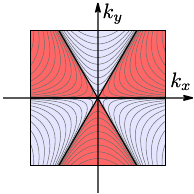}
\end{tabular}\\
\begin{tabular}{@{}c@{}}
Parabolic\\
umbilic\\
($D_5$)
\end{tabular}
&
$(2,4,4,4)$
&
$|\epsilon|^{-3/8}$
&
$\frac{1}{2} \csc \left(\frac{\pi }{8}\right)$
&
\begin{tabular}{@{}c@{}}
$k_x^2 k_y+k_y^4 $ \\
$+ t_4 k_x^2 + t_3 k_y^2 + t_2 k_x + t_1 k_y$
\end{tabular}
&
\begin{tabular}{@{}c@{}}
\includegraphics[scale=0.5]{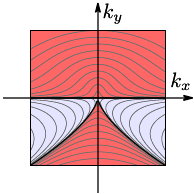}
\end{tabular}\\
\begin{tabular}{@{}c@{}}
$X_9$ \\
(for $c = -3$)
\end{tabular}
&
\begin{tabular}{@{}c@{}}
$(2,8,4,8)$\\
\end{tabular}
&
$|\epsilon|^{-1/2}$
&
\begin{tabular}{@{}c@{}}
$1$\\
\end{tabular}
&
\begin{tabular}{@{}c@{}}
$k_x^4 + k_y^4$ \\
$+2ck_x^2 k_y^2 + t (k_x^2 + k_y^2)$\\
(unfolding consistent \\with $\pi/2$-rotation)
\end{tabular}
&
\begin{tabular}{@{}c@{}}
\includegraphics[scale=0.5]{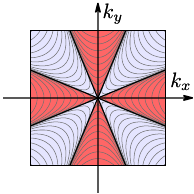}
\end{tabular}\\
\hline
\end{tabularx}
\end{center}
\end{table}

\subsection{Examples in 1D and 3D}
In 1D let us consider the dispersion relation $E=A_n k^n$. Then, the topological index is $I= \frac{1}{2} \left[1 + (-1)^n\right] sgn(A_n)$ \cite{LiangFu}. 
Since the dispersion is bounded, it will at least have a minimum 
$I=-1$ 
and a maximum 
$I=+1$, 
which already satisfies the topological
constraint 
$\sum_i I_i=0$. As a result, 
the dispersion does not necessarily exhibit a saddle point $I=0$ in 1D (in contrast to the 2D case, where $I=+1$ for local extrema and $I\leq 0$ for saddle points). 
The 1D DOS diverges in a power-law fashion already for an ordinary critical point $\rho(\epsilon)\sim \epsilon^{-\gamma}$ with $\gamma=1/n-1$. 

In three dimensions, we can consider an ordinary critical
point with dispersion $\epsilon({\bf k}) = {\bf k}^T D {\bf k}$ with
Hessian matrix $D$. Then, the topological index is
$I = sgn(detD)$ 
and a local minimum (maximum) has topological
index $I = +1$ ($I = -1$), 
while 
for a saddle point, the topological index can be either +1
or -1 \cite{LiangFu}. 
More generally, the topological index of an isolated critical
point in 3D is 
given by 
the number $n_e$ $(n_h)$ of 
energy surfaces with positive $E > 0$ (negative $E < 0$) energy in the region enclosed by $\mathcal{S}$: $I = n_e-n_h$ \cite{LiangFu}. 
This implies that the topological index in 3D is odd under energy inversion $E \rightarrow -E$, which exchanges $n_e\longleftrightarrow n_h$. 
Thus, it is also possible in 3D, albeit tuning is required, that the dispersion does not possess a saddle point because the topological constraint can be satisfied by a minimum and a maximum. The DOS exponent $\gamma=\sum_i a_i -1$ in 3D is not necessarily negative so that the DOS can be finite.  

\subsection{Response} 

HOVHS are associated with changes in the topology of the Fermi surfaces and a non-analyticity in the DOS. 
The response to these changes can be a signal for the occurrence of a HOVHS.  We take the example of a 2D interacting electrons system at a monkey saddle of biased bilayer graphene with dispersion relation $E\propto k_x^3 - 3 k_xk_y^2$ which arises when three Van Hove saddles merge in an elliptic umbilic elementary catastrophe \cite{Shtyk_2017} (see Tab.~\ref{table1}). A HOVHS of this kind can be identified,  
e.g., 
by its distinct behavior of the energy of the $m^{th}$ Landau level $E_m$ as a function of an applied magnetic field $B$ with $E_m \propto (Bm)^{3/2}$ \cite{Shtyk_2017}.  There are also related oscillations in thermodynamic and transport properties, such as de Haas–Van Alphen and Shubnikov–de Haas oscillations, whose period triples as the system crosses the singularity.
This behavior can be generalised to analogous types of HOVHS where $n$ Fermi surfaces merge such as $X_9$ in Tab~\ref{table1} to $E_m \propto (Bm)^{n/2}$ in a semiclassical approximation.
The period when $n$ pockets are disconnected is $n$ times the one of the single Fermi surface when they are connected \cite{Shtyk}.

More evidence for a Lifshitz transition in general, and connected to a HOVHS in particular, can be obtained from other magneto-transport properties. For example, the Hall density 
strongly increases before changing sign at Van Hove filling because the carriers change between electron- and hole-like. 
A semiclassical theory for the magneto-transport properties of metals with Fermi surface topological transitions in 2D is developed in Ref.~\cite{PhysRevB.105.155123}. Signatures of Fermi surface topological transitions in general, in the Hall coefficient, Hall conductivity $\sigma_{xy}$, and longitudinal conductivity $\sigma_{xx}$ are provided as a function of both a perpendicular magnetic field and a time-dependent electric field. The possibility of a nonzero Berry curvature was taken into account.  
The analysis is valid away from Fermi surface topological transitions. Very close to them 
one has to consider the possibility of magnetic breakdown. The reason for this magnetic breakdown is that two or more semiclassical cyclotron orbits approach each other and quantum tunneling between several trajectories may happen. Then, nontrivial scattering amplitudes and phases are obtained as shown in Ref. \cite{PhysRevB.109.L081103} by mapping the problem to a scattering problem in a 1D tight-binding chain. The quantum tunneling facilitates the delocalization of the bulk Landau level states and the formation of 2D orbit networks. The effect can be observed in transport experiments as a strong enhancement of the longitudinal bulk conductance in a quantum Hall bar \cite{PhysRevB.109.L081103}. Other signatures of a power-law DOS in transport include a non-Fermi-liquid scaling in the temperature dependence of the dc resistivity \cite{Zang2022,Mousatov2020}, which deviates from the one of ordinary VHS \cite{Hlubina1996,Buhmann2013,Herman2019}. 
It is worth mentioning that in the presence of low concentration and short-range impurities, although the divergence of the DOS is smeared, the shape of the DOS, as characterized by the power law tail and the universal ratio of prefactors, is retained slightly away from the singularity \cite{Chandrasekaran_Betouras_2022}.

\section{INTERACTION EFFECTS}

In a flat-band limit, the coupling between electrons is larger than the kinetic energy and the band dispersion can be considered negligible or a perturbative correction. However, this situation can change when bands are only partially flat, e.g., due to approximate symmetries (see Sec.~\ref{sec:flatbands}). In addition, effective interactions for the states within the narrow bands can be screened from other high-energy bands and acquire a non-trivial momentum dependence from projection into the narrow bands. Then, when the bandwidth becomes larger than the average coupling strength, it is advantageous to consider the band structure as a starting point and include interactions perturbatively. In this case, VHS are of particular importance when it comes to interaction effects. 
The role of a VHS in the DOS was discussed for many correlated materials because the singular DOS suggests that an instability of the Fermi liquid arises at low temperature. Interaction-corrections to the Fermi liquid are determined by the dimensionless coupling $g \rho(\epsilon_F)$, which is the product of coupling strength $g$ and DOS at the Fermi level $\rho(\epsilon_F)$. These corrections 
diverge with the distance of the Fermi level to the VHS, signaling a breakdown of perturbation theory. Generically, this can happen in different scattering channels so that a competition of orders is expected.  In particular, a pairing instability also becomes possible because strong spin or charge fluctuations provide a pairing glue. 
This Van Hove scenario is qualitatively changed at a HOVHS in comparison to the conventional VHS. 
For a single HOVHS in the BZ and weak repulsive interaction, a non-Fermi-liquid state that was coined supermetal develops instead of an ordered state \cite{Isobe_Fu}. For attractive interaction, there is a pairing instability, but with an increased, non-BCS critical temperature. In case there are several HOVHS in the BZ, the supermetal becomes unstable and instabilities towards different orders compete \cite{HOvHS2}. However, the HOVHS can drive different types of orders compared to the conventional case. For example, Pomeranchuck instabilities or pair density waves can become competitive compared to spin or charge density waves and zero-momentum pairing \cite{Wu2023PDW}.

\subsection{Bare susceptibilities}
The breakdown of perturbation theory manifests itself in a divergence of the static susceptibilities in the particle-particle (pp) and particle-hole (ph) channel. They are defined as 
\begin{align}\label{eq:bubbles}
    \Pi_{pp,ph}(\vec q)=\pm T \sum_{i\omega_n} \int \frac{d \vec k}{A_{BZ}} G(i\omega_n, \vec k)G(\mp i\omega_n, \mp\vec k+\vec q)\,,
\end{align}
where $G=(i\omega_n-\xi_{\vec k})^{-1}$ is the free electron Green's function, the sum goes over Matsubara frequencies $\omega_n=(2n+1)\pi T$ and the integral over the area of the Brillouin zone $A_{BZ}$. 
At a conventional VHS $\Pi_{pp}(0)\sim \ln^2 T$ and $\Pi_{ph}(0)\sim \ln T$. Additionally, there can be a singularity at the wave vector $q=Q$ that connects different Van Hove points, where $\Pi_{pp}(Q)\sim \ln T$ and either $\Pi_{ph}(Q)\sim \ln T$ or $\Pi_{ph}(Q)\sim \ln^2 T$ with or without nesting $\xi_{\vec k+\vec Q}=-\xi_{\vec k}$. 
For the HOVHS case, we write the DOS as $\rho(\omega)=D_\pm \omega^{-\gamma}$, where both $D_\pm$ and $\gamma$ depend on the type of HOVHS (see Sec.~\ref{sec:classes}). Some pp and ph susceptibilities inherit this power-law dependence. For zero wave vector transfer, this happens in both channels 
\begin{equation}
\Pi_{pp}(0)=\rho_0 f_1(\gamma) T^{-\gamma} \;\;\;\;\;\;
\Pi_{ph}(0)=\rho_0 f_2(\gamma)T^{-\gamma}\,,
\end{equation}
where $\rho_0=(D_+ + D_-)/2$, $f_1(\gamma)=(1/2)\int d\epsilon |\epsilon|^{-(1+\gamma)}|\tanh(\epsilon/2)|$ and $f_2(\gamma)=(1/4)\int d\epsilon|\epsilon|^{-\gamma}\cosh^{-2}(\epsilon/2)$ with $\Pi_{ph}(0)/\Pi_{pp}(0)= \gamma$. 
Note that the order of limits is important to obtain a non-zero result for $\Pi_{ph}(0)$ when $T\rightarrow 0$ due to the double-pole structure. 
In the low-energy limit, the Fermi surface becomes scale invariant and the theory super-renormalisable, i.e. free of UV divergencies. This implies that only one energy cutoff is sufficient to regularise Eq.~\eqref{eq:bubbles}, in contrast to the case of an ordinary VHS, where a second momentum cutoff is needed.
For finite wave vector transfer $Q$, the susceptibilities can but don't need to diverge with a power-law $\Pi_{pp,ph}(Q)\sim T^{-\gamma}$, depending on details of the dispersion such as the presence of nesting. 
As we can see from these expressions at an ordinary VHS without nesting, the divergence in pp channel is parametrically larger than in ph channel ($\ln^2T$ vs $\ln T$). In contrast, at a HOVHS, singularities in pp and ph channel are of the same form $T^{-\gamma}$ so that they necessarily compete \cite{Zervou2023}. 

\subsection{Ladder series}
The divergences of the static susceptibilities imply spin, charge, or pairing instabilities when the corresponding ladder series are summed up separately (see below why this is not necessarily a good approximation). Introducing a test field $\Gamma^0_{l, q}$ for an instability with characteristic wave vector $\vec q$ in angular-momentum pp or ph channel $l$, the ladder series results in a correction of the form
\begin{align}
\Gamma_{l,q}=\frac{\Gamma^0_{l, q}}{1-g_{l,q} \Pi_{l}(\vec q)}\,,
\end{align}
where $g_{l,q}\lessgtr 0$ denotes the coupling in the corresponding channel. An instability arises if $1-g_{l,q }\Pi_{l}(\vec q)=0$. For a power-law divergent susceptibility $\Pi_l(\vec q)\propto T^{-\gamma}$ so that the corresponding critical temperature scales in a non-BCS fashion 
\begin{align}
T_c\propto g_{l,q}^{1/\gamma}\,,
\end{align}
and is significantly enhanced.
For example, the Cooper ladder, i.e. the s-wave pairing channel, possesses such an instability if the bare interaction $g_0$ is attractive $g_{s,pp}=-g_0>0$.  

\subsection{Competing orders}
The ladder approximation is justified when the corrections in one channel $g_{l,q}\Pi_l(\vec q)$ are much larger than in all others. However, around HOVHS, corrections of ph and pp channels are of the same order, and we cannot neglect their coupling. Both pure ladder diagrams as well as mixed-channel diagrams are of the order $T^{n\gamma}$ on the $n$-loop level. Thus,  it is crucial to account for the feedback of fluctuations of one channel on another. 
This can be achieved via a renormalisation group (RG) approach that keeps all leading divergencies on the same footing. Such an RG approach is typically performed in an effective, low-energy theory for the states in the Brillouin zone that are close to the HOVH points and subject to all symmetry-allowed interactions between them. It is argued that it is sufficient to consider only these states because they are the ones that govern thermodynamic contributions due to their singular DOS. 
For example, in the case of the triangular lattice that we considered in Sec.~\ref{sec:tuningtoHOVHS}, the effective, low-energy theory becomes $H=H_0+H_g$, where $H_0$ contains the dispersion in a patch of size $\Lambda$ around HOVH points and $H_g$ all symmetry-allowed interactions between states near HOVH points. 
Explicitly, they are given by 
$H_0=\sum_{p,k,\sigma}\epsilon_{p,k} c_{p\sigma k}^\dagger c_{p\sigma k}$ with the dispersion around  $M_p$, $p\in \{1,2,3\}$, up to quadratic order $\epsilon_{p,k}=\epsilon(M_p+k)$ as in Eq.~\eqref{eq:HOdispM1} and 
\begin{align}
H_{g}=\sum_{\substack{k_1\ldots k_3\\\sigma \sigma'}}\sum_{\substack{p,p'=1\\p\neq p'}}^3 \Big[&\ g_1 c_{p'\sigma k_3}^\dagger c_{p \sigma k_4}^\dagger c_{p' \sigma' k_2} c_{p \sigma k_1} 
+  g_2 c_{p \sigma k_3}^\dagger c_{p' \sigma k_4}^\dagger c_{p' \sigma' k_2} c_{p \sigma k_1}\nonumber \\
+ &\ g_3 c_{p' \sigma k_3}^\dagger c_{p' \sigma k_4}^\dagger c_{p \sigma' k_2} c_{p \sigma k_1}\Big] 
+\sum_{\substack{k_1\ldots k_3\\\sigma \sigma'}}\sum_{\substack{p=1}}^3
\ g_4  c_{p \sigma k_3}^\dagger c_{p \sigma k_4}^\dagger c_{p \sigma' k_2} c_{p \sigma k_1}
\end{align}
with four couplings $g_i$, $i\in \{1,...,4\}$ \cite{HOvHS1}\footnote{The low-energy patch model for honeycomb and triangular lattice is equivalent.}. The operators $c_{p\sigma k}$ annihilate states in the vicinity of $M_p$ with momentum $M_p+k$ and spin $\sigma$. 
The RG equations sum up all possible vertex corrections of order $T^{n\gamma}$ as ph and pp corrections are taken into account on the same level. This gives \cite{HOvHS1}
\begin{align}
	\dot{g}_1&=\dot{\Pi}^0_{\mathrm{ph}}\left[(N-2)g_1^2+2g_1g_4\right]\nonumber\\
    \dot{g}_2&=\dot{\Pi}^0_{\mathrm{ph}}\left[2g_4(g_1+(1-N_f)g_2)+(N-2)g_2(2g_1-N_fg_2)\right]\nonumber\\
	\dot{g}_3&=-\dot{\Pi}^0_{\mathrm{pp}}\left[2g_3g_4+(N-2)g_3^2\right]\nonumber\\
	\dot{g}_4&=\dot{\Pi}^0_{\mathrm{ph}}\left[(N-1)(g_1^2+2g_1g_2-N_fg_2^2)+(3-N_f)g_4^2\right]-\dot{\Pi}^0_{\mathrm{pp}}\left[g_4^2+(N-1)g_3^2\right]
\end{align}
where $\dot{g}_i\equiv\frac{d}{dt} g_i$ and $\dot{\Pi}_{pp/ph}^0 \equiv \frac{d}{dt} \Pi_{pp/ph} (0)$ and $t=\ln \Lambda/T$. These equations are valid for general flavour number $N_f$ ($N_f=2$ for spin 1/2) and number of patches $N$ for square ($N=2$) or triangular and honeycomb ($N=3$) lattice. 
The RG analysis is formally controlled by the small exponent $\gamma$; it approximates the mixed contributions by the product of decoupled ph and pp diagrams, which is justified to logarithmic accuracy in the ordinary case $\gamma\rightarrow 0$.
Generally, the RG procedure finds a flow to strong coupling, i.e., the renormalised couplings diverge for decreasing temperature $T$, which is again a manifestation of the breakdown of perturbation theory and signals the formation of an ordered state. However, the divergence occurs in a meaningful way, in the sense that it singles out a certain interaction channel to diverge strongest and the ordering tendency can be identified by the corresponding leading susceptibility. This can be checked in a similar way as for the ladder series by introducing test fields $\hat\Gamma_{l,q}$, which are now scale-dependent and receive corrections from the renormalised couplings. 
In the case of the triangular lattice, instabilities towards ferromagnetism, d-wave pairing, or nematic (d-wave Pomeranchuk) order occur for repulsive initial interactions \cite{HOvHS1}.
Such an RG procedure was performed for many materials and model systems. They include doped graphene \cite{HOvHS1}, Bernal bilayer graphene \cite{Shtyk_2017}, twisted bilayer graphene \cite{HOvHS2, Sherkunov-Betouras, PhysRevX.8.041041}, twisted trilayer graphene \cite{HOvHS4}, twisted WSe${}_2$ \cite{Wu2023PDW,Hsu2021,Wu2023PRL,Klebl2023}, cuprates \cite{Markiewicz2023}, or kagome metals \cite{Han2023}. In model systems, e.g., Rashba spin orbit coupling \cite{HAN2024319}, topological bands \cite{Castro2023,Aksoy2023}, or 3D lattices \cite{Igoshev2023} were investigated. Furthermore, the interplay between ordinary and HOVHS \cite{Zervou2023,lee2024crossover} was studied along similar lines. 

\subsection{Supermetal}

Assuming a single HOVH point is present in the BZ with an energy close to the Fermi level, the corresponding effective, low-energy theory contains just one contact interaction $(g/2) (\psi^\dagger\psi)^2$ and power-law divergent susceptibilities arise only for zero wave vector. Interestingly, for a weak, repulsive bare interaction $g>0$, these instabilities do not grow into an ordered ground state. Instead, the power-law corrections drive the Fermi liquid into a critical non-Fermi-liquid state. This can be seen at the RG equations which admit another type of solution unique to the HOVH case:  an interacting fixed point for the dimensionless coupling $\hat g=g \rho(\Lambda)$, where $\Lambda$ is the RG cutoff scale $\Lambda$ \cite{Isobe_Fu,Shtyk_2017}. 
While this type of solution is typically unstable when more than one HOVH point is present in the BZ, it becomes the stable solution for just one HOVHS point. The corresponding fixed point is formally analogous to the Wilson-Fisher fixed point of (bosonic) $\phi^4$ theory. As a result, the coupling does not diverge in the infrared, but flows to a finite value. 
This describes a ground state - coined supermetal \cite{Isobe_Fu} - which remains metallic without long-range order but with power-law divergent charge and spin susceptibilities right at criticality. A two-loop analysis shows that the electrons also acquire a finite anomalous dimension 

\subsection{Interaction-induced band renormalisation} 
\label{sec:selfenergy}
Interactions also affect the single-particle band structure via self-energy corrections. They play a role for HOVHS in two different ways. On the one hand, HOVHS represent a fine-tuned situation in which certain curvature terms of the dispersion are zero. Since these terms are not symmetry-forbidden, self-energy corrections are generically expected to re-generate them and, thus, spoil the high-order behavior. Yet, at least in a weak coupling approach, self-energy corrections can be estimated to be subleading to scattering corrections because they appear at two-loop order. The first non-analytic contribution to the self-energy scales as $\Sigma\sim T^{1-2\gamma}$ compared to $\Pi\sim T^{-\gamma}$ of the one-loop susceptibilities \cite{Isobe_Fu,HOvHS1}. Therefore, it is possible that an ordering instability sets in at a higher temperature before self-energy correction become relevant. 

On the other hand, it is, in principle, possible that self-energy corrections produce the opposite effect. Starting with an ordinary Van Hove point, it may be energetically advantageous to pin the Van Hove point to the Fermi level and flatten the bands towards a HOVH point so that interactions become more effective in gapping out the region of large DOS. This would be consistent with the observation of extended flat dispersions in ARPES data of several correlated metals \cite{Rosenzweig_2020,Hu2022,Link2019,evHs1,evHs3,evHs4,evHs5,Lu_Bednorz_1996,Yokoya_Tokura_1996}. Theoretically, the tendency towards interaction-induced band flattening was demonstrated for square and triangular lattices \cite{Irkhin2002,Yudin2014}, as well as for a more general class of 2D models with a non-nested Fermi surface that contains a Van Hove point \cite{FeldmanSalmhofer2008}.

\section{MATERIALS}

\subsection{Strontium Ruthenates}

\begin{figure}[t]
	\center{\includegraphics[width=1.\linewidth]{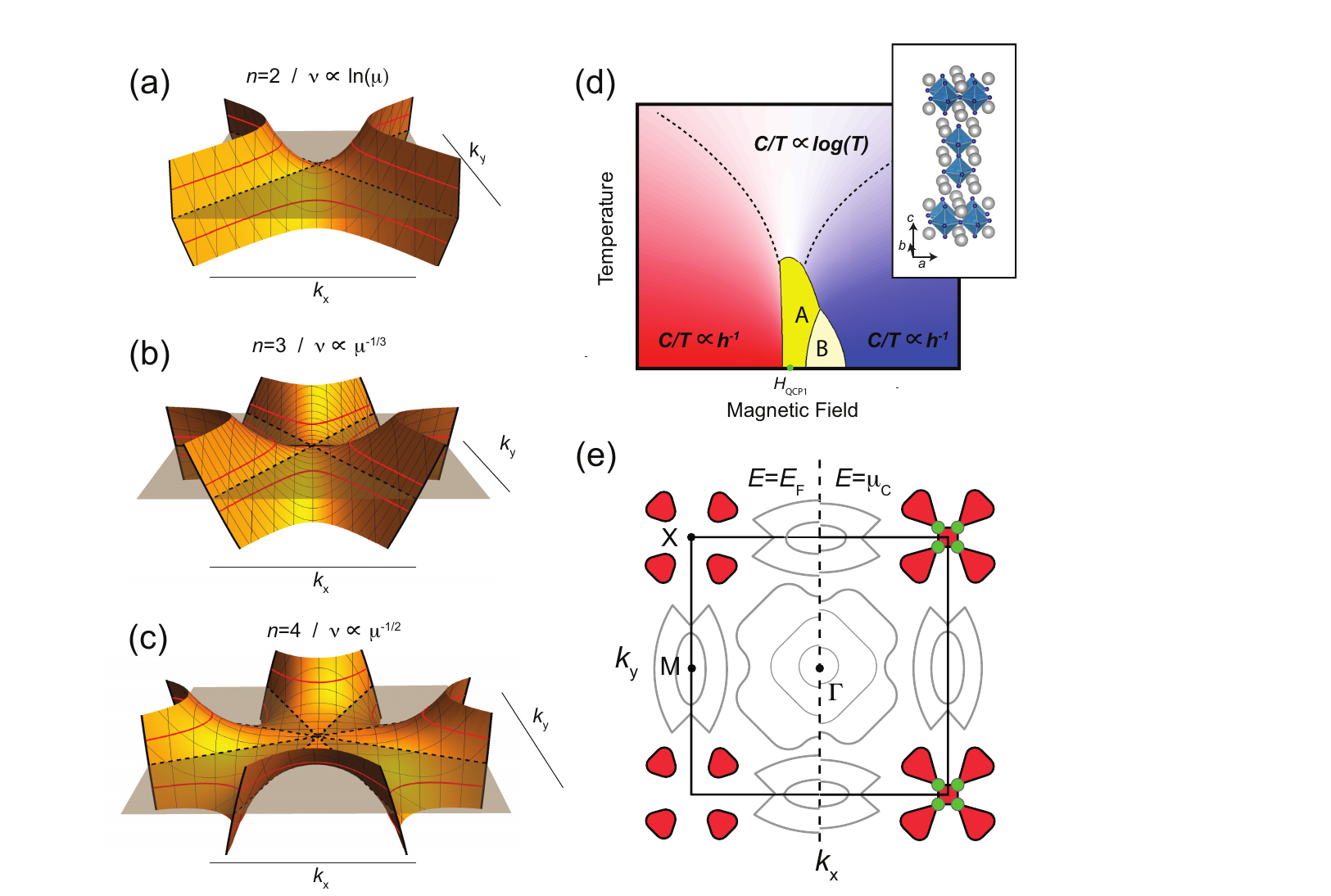}} 
	\caption{(a) The $n$=2 2D regular VHS in the form of a saddle point. The DOS $\nu$ diverges as $\ln\mu$. Red lines indicate Fermi surfaces above and below the singularity. The critical Fermi surface is shown by the dotted black line.  (b) The $n$=3 singularity that can occur at 3-fold symmetric points. (c) The $n=4$ singularity at a 4-fold symmetric point, it occurs very close to the Fermi energy in the case of Sr$_3$Ru$_2$O$_7$.  (d) The thermodynamic phase diagram of Sr$_3$Ru$_2$O$_7$ where the different regions (red, blue, mixed) correspond to different behavior of the specific heat C/T as a function of an applied magnetic field. 
    The regions A and B correspond to two spin-density-wave phases. (e) Schematic of the quasi-2D FS of Sr$_3$Ru$_2$O$_7$ in the $k_z$=0 plane at the Fermi energy (left hand side) and at $\mu_\textrm{C}$ (right side). The crucial bands that are close to the $n=4$ multicritical point are highlighted in red. The central pocket is a small perturbation. In order to emphasise the characteristic clover leaf Fermi surface we show an extended $k$-space picture beyond the BZ boundaries. Figure adapted from \cite{Efremov_PRL_2019}.}
	\label{fig:structure}
\end{figure}

The Ruddlesden–Popper (RP) series of strontium ruthenates (Sr$_{n+1}$Ru$_n$O$_{3n+1}$, $n$ = 1, 2,···) have attracted intensive research interest as model systems of correlated quantum materials, mainly due to their relatively simple structure that at least the two first members of the family possess. These systems are among the first systems that offer experimentally accessible HOVHS with consequences on 
observables.  

The first compound of the series with compelling evidence to exhibit HOVHS is the bilayer Sr$_3$Ru$_2$O$_7$. This material displays anomalous field-dependent electronic properties and quantum criticality which had been the subject of debates. Early proposals suggested that critical fluctuations were associated with metamagnetism or nematicity. The discovery of spin density wave order in Sr$_3$Ru$_2$O$_7$ 
questioned these proposals while earlier ARPES and DFT results showed evidence of a VHS near the Fermi energy \cite{Tamai_PRL_2008}. Sr$_3$Ru$_2$O$_7$ is magnetically ordered within a magnetic field over a small range of field values. This effect was shown to be a consequence of a Fermi surface topological transition that leads to a HOVHS of 
X$_9$ type (energy dispersion proportional to k$^4$ cos($4\phi$) where $\phi$ is the azimuthal angle, see Tab.~\ref{table1}) and DOS that diverges $\rho(\epsilon) \propto \epsilon^{-1/4}$, see Fig.~\ref{fig:structure}.
A unique experimental signature of this dispersion relation is the magnetic field-dependence of the specific heat which is $C/T \propto 1/|H-H_c|$, where $H_c$ is the critical value of the applied field that the HOVHS emerges at the Fermi energy \cite{Efremov_PRL_2019}. The changes of the Fermi surface topological transition affect the nesting and the wavevector-dependent susceptibility $\chi(Q)$ \cite{Efremov_PRL_2019, Lester_NatMat_2015, Lester_NatComms_2021}.

Another well studied ruthenate is Sr$_2$RuO$_4$, a material known to superconduct with a structure very similar to the high-temperature cuprate superconductors.  There has been a long-standing question on the appearance of HOVHs in Sr$_2$RuO$_4$, as evidenced through ARPES measurements very early \cite{Lu_Bednorz_1996, Yokoya_Tokura_1996}, before surface and bulk contributions were distinguished \cite{Damascelli2000}. In the bulk, an ordinary VHS is available close to Fermi energy, and upon tuning the system through uniaxial stress \cite{Sunko2019} or strain, the superconducting T$_c$ is significantly boosted \cite{Hicks_Science_2014, Steppke2017}. 
On the other hand, the surface exhibits a reconstruction due to octahedral rotations \cite{Damascelli2000, Matzdorf2002, Marques2021, Morales_PRL_2023} providing access to the electronic structure in parameter regimes not accessible in the bulk. As a result, the surface layer of Sr$_2$RuO$_4$ provides an ideal test system to explore the symmetries of VHS in the electronic structure: the two-dimensional nature of the electronic structure, evidenced by de Haas-van Alphen measurements and a resistivity anisotropy of more than 500 \cite{mackenzie_quantum_1996}, means that it can be fully captured using surface spectroscopies \cite{Damascelli2000, Marques2021, Tamai2019, wang_quasiparticle_2017}. 
In particular, the octahedral rotations in the surface layer of Sr$_2$RuO$_4$ mean that the $d_{xy}$ VHS, which is at the zone face in the bulk of Sr$_2$RuO$_4$, is at the zone corner in the surface layer, conducive to the formation of a $X_9$ HOVHS as in Sr$_3$Ru$_2$O$_7$ \cite{chandrasekaran2023engineering}. Taken as an example, this system also served for further studies on how to manipulate and engineer HOVHS through strain \cite{chandrasekaran2023engineering}.

Recently, the trilayer Sr$_4$Ru$_3$O$_{10}$ has been reported to exhibit multiple Van Hove singularities (still to be determined whether HOVHS) in the vicinity of the Fermi level. The study of the spectral function measured by ARPES and quasiparticle interference reveals that the VHS closest to the Fermi energy emerges due to a combined influence of octahedral rotations and spin-orbit coupling \cite{marques2023spinorbit}.

\subsection{Kagom\'e metals}

Another class of materials with intense research interest and evidence of harbouring HOVHS is the kagom\'e metals. Especially, the layered vanadium antimonides AV$_3$Sb$_5$ (A=K, Rb, Cs) are a recently discovered family of topological kagom\'e metals that exhibit a range of strongly correlated electronic phases including charge order and superconductivity. 

The kagom\'e lattice, as already explained in Sec.~\ref{sec:flatbands}, has a characteristic electronic structure that features flat bands across the whole Brillouin zone, in addition to Dirac fermions at the Brillouin zone corner K and VHS at the zone edge M. When spin-orbit coupling is included both the quadratic band touching point at the zone centre $\Gamma$ and the linear band crossing at the Brillouin zone corner K are gapped \cite{Guo_PRB_2009, Tang_PRL_2011, Xu_PRL_2015}. A main quest is to discover compounds that realise specific properties of the kagom\'e lattice.
A series of 3d transition-metal-based kagom\'e compounds which have been intensively studied, realize 
the topological physics of the lattice. Examples include  FeSn \cite{Kang_NatMat_2020} and CoSn \cite{Kang_NatComms_2020} that show topological flat bands \cite{jiang2023kagome},  Fe$_3$Sn$_2$ \cite{Ye_Nature_2018}and TbMn$_6$Sn$_6$ \cite{Yin_Nature_2020} which realize massive Dirac fermions and intrinsic anomalous Hall conductivity while in  Mn$_3$Sn, Mn$_3$Ge and Co$_3$Sn$_2$S$_2$  magnetic Weyl fermions and chiral anomaly were realized \cite{Nakatsuji_Nature_2015, Nayak_SciAdv_2016, Liu_NatPhys_2018, Liu_Science_2019}.

AV$_3$Sb$_5$ (A=K, Rb, Cs) display a number of correlated ground states in their phase diagram 
which were related to VHS near the Fermi level. These ground states include a 2 $\times$ 2 inverse star-of-David charge order below T$_{CO}$ $\approx$ 78–102 K, and superconductivity below T$_c$ = 0.92–2.5K \cite{Ortiz_PRM_2019, Ortiz_PRL_2020, Ortiz_PRM_2021, Yin_CPL_2021}.  
Josephson microscopy and thermal conductivity experiments indicated that the superconductivity is in the strong coupling regime $(2\Delta_{sc}/k_B T_c\approx 5.2)$ a \cite{Chen_Nature_2021}. High-pressure measurements showed the competition between the charge order and superconductivity with a double superconducting dome in the temperature-pressure phase diagram. Moreover, nematicity that breaks the four-fold rotational symmetry  and a four-unit-cell periodic stripe phase were also discovered below 60K in coexistence with superconductivity.

\begin{figure}[t]
	\center{\includegraphics[width=1.0\linewidth]{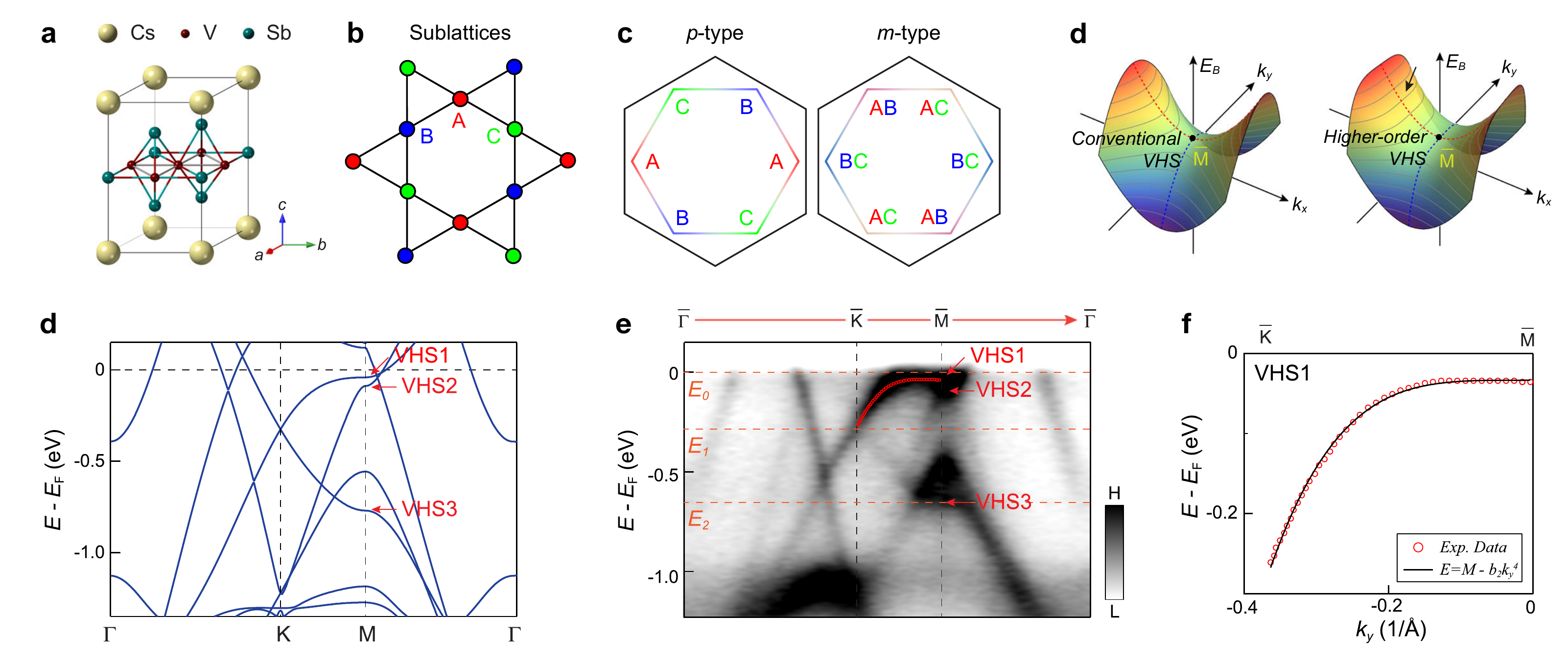}} 
	\caption{(a) The lattice structure of kagome metals CsV$_3$Sb$_5$. (b) Real space structure of the kagome vanadium planes; red, blue, and green coloring indicate the three sublattices. (c) Two distinct types of sublattice contributions to van Hove points in CsV$_3$Sb$_5$: p-type (sublattice pure, left panel) and m-type (sublattice mixing, right panel). (d) Schematics of dispersion around ordinary and HOVHS. The gray curves indicate constant energy contours that show markedly flat features along the $k_y$ direction in the HOVHS case as highlighted by the black arrow. (e) Experimental band dispersion along $\Gamma-K-M-Gamma$ direction. Dashed lines indicate the energy position of the Fermi level, Dirac cone and VHS$_3$. (f) Calculated bands along $\Gamma-K-M-Gamma$ direction. The red arrows in (e, f) mark multiple VHS. (g) Fitting to the measured dispersion along $M-K$ by $E=M-b_2 k_y^4$ (solid black line). The red dots represent the experimental data shown in (e).  Adapted from Ref.~\cite{Hu2022}.}
	\label{fig:KagomeHOVHS}
\end{figure}

A combination of ARPES with DFT studies revealed and refined the role of multiple VHS coexisting near the Fermi level of CsV$_3$Sb$_5$ \cite{Kang2022,Hu2022}, see Fig.~\ref{fig:KagomeHOVHS}. The VHS are characterized by two distinct sublattice flavours (p-type and m-type), which originate from their pure and mixed sublattice characters, respectively. These twofold VHS flavours of the kagom\'e lattice determine the pairing symmetry and unconventional ground states \cite{Kiesel2012}. 
Four VHSs were identified in CsV$_3$Sb$_5$ around the M point, three of them close to the Fermi level, with two having sublattice-pure and one sublattice-mixed nature \cite{Hu2022}. More specifically the m-type VHS of the d$_{xz}$/d$_{yz}$ kagom\'e band and the p-type VHS of the d$_{xy}$/d$_{x^2–y^2}$ kagom\'e band are located very close to the Fermi level \cite{Kang2022}. The m-type VHS band is characterized by pronounced Fermi surface nesting, while the p-type displays an extremely flat dispersion along MK, establishing the experimental discovery of a HOVHS \cite{Kang2022, Hu2022}. These properties seem to hold for all members of the AV$_3$Sb$_5$ series.

Recently the magnetic kagom\'e compound GdV$_6$Sn$_6$ was also found to display VHS near the Fermi energy along with tunable topological Dirac surface states \cite{YongHu2022}.

\subsection{Graphene systems and moiré materials}
\label{sec:graphenemat}

In the comparatively simple band structure of graphene VHS can be found at high energies in the charge-neutral system requiring significant doping to bring them to the Fermi level. Such doping levels were successfully achieved using intercalation combined with additional adsoprtion. ARPES measurements on the doped system revealed an extreme flattening of the dispersion around the VHS providing evidence for the formation of a HOVHS \cite{Rosenzweig_2020,evHs1,Link2019}. The significant band renormalisation was interpreted as a correlation effect due to electronic interactions, as well as strong electron-phonon coupling. 

\begin{figure}[t]
	\center{\includegraphics[width=.99\linewidth]
 {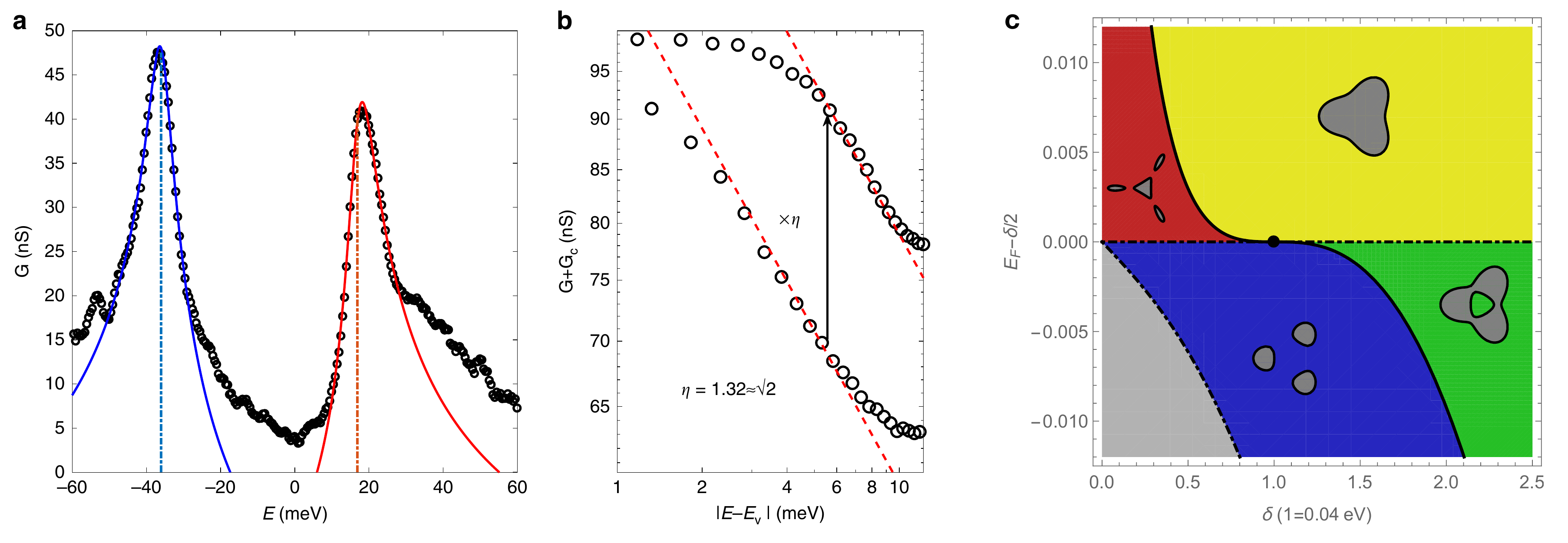}} 
	\caption{(a) Measured tunneling conductance $G$ (open circles) of twisted bilayer graphene with angle $1.10^\circ$ and theoretical fit (solid lines). Dashed lines indicate peak positions. (b) Conductance minus a background offset $G_c=57.6$ nS around the peak at $E_V=16.72$ meV on a logarithmic scale. Red dashed lines correspond to power-law scaling $|E-E_V|^{-1/4}$ on both sides of the peak with asymmetric ratio $\eta$. Figures 4(a) and 4(b) are taken from Ref.~\cite{Yuan2019}. (c) Fermi-surface topologies in biased bilayer graphene as function of displacement field $\delta$ and Fermi level $E_F$. They are separated by band-edge (dash-dotted) or ordinary Van Hove transitions (solid). There is a HOVH point at the crossing of these lines. In the gray area, the Fermi level lies in the band gap. Taken from Ref.~\cite{Shtyk_2017}.}
	\label{fig:TBGHOVHS}
\end{figure}

In Bernal bilayer graphene, another type of HOVHS becomes accessible without chemical doping. Its low-energy band structure possesses four Dirac cones and three Van Hove points per valley due to trigonal warping and extended hopping terms. This complex band structure can be manipulated by an electric displacement field between layers, which is independently controllable from the carrier density using top and bottom gate voltages. The displacement field introduces a gap and flattens the top (bottom) of the valence (conduction) band. It also moves the location of the three VHS which merge into a HOVHS at a critical value of the field (classified as "elliptic umbilic" in Tab.~\ref{table1}), see Fig.~\ref{fig:TBGHOVHS}. Changes in the Fermi surface topology and the tunability of VHS were recently demonstrated in a detailed analysis of the Landau level spectrum \cite{seiler2023probing,Varlet2014}. The complex evolution of the band structure is also of high relevance for the reported rich many-body phase diagram which shows a cascade of symmetry-broken ground states accompanied with non-trivial Fermi surface reconstructions \cite{seiler2022, Zhou_2021,lin2023spontaneous}. 

VHS can also be brought in the vicinity of the Fermi level in moir\'e materials, e.g., reported in \cite{Yuan2019,WuAndrei2021,Park2021,Wang2020}. Owing to their high tunability, 
theory further often predicts HOVHS in the accessible parameter regime \cite{Wang2021,HOvHS4}, including in twisted bilayer graphene \cite{Yuan2019}. In the latter case, strong direct evidence was provided in the tunneling conductance, see Fig.~\ref{fig:TBGHOVHS}.  It shows a power-law scaling in a range of energies around the conductance peaks. The extracted power law as well as the universal asymmetry ratio of the two sides of the peak $D_+/D_-$ are consistent with the expected scaling of a cusp HOVHS (see Tab.~\ref{table1}).

Evidence for a tunable VHS with displacement field were also reported for twisted bilayer WSe${}_2$ \cite{Wang2020}. The sign change of the Hall resistance evolves continuously with doping and displacement field, consistent with theoretical modelling of the VHS position. When the VHS is near half filling of the moir\'e band, a resistance peak was observed indicating the formation of a correlated insulator. The magnitude of the peak correlates closely with the distance of the VHS to the Fermi level. The VHS that occurs at half filling in the theoretical models is of high order \cite{Pan2020,Zang2021,Zang2022} suggesting a close connection to the observed behavior.

\section{OVERVIEW AND OTHER POSSIBLE MATERIALS}

In the beginning of this review we outlined strategies that theoretically lead to the appearance of HOVHS, before we presented the evidence for their presence in selected quantum materials in the previous section. One of these strategies 
draws a close connection to flat bands.  
If materials crystallized in lattices that, in principle, exhibit flat bands, these bands acquire dispersion 
in realistic systems because of perturbations from the ideal models (approximate symmetries, further-range hopping, etc). 
Then HOVHS can appear as a remnant of the exactly flat band. 
A systematic analysis how a spectrum gets flattened from locally (Van Hove) to globally flat dispersions 
promises valuable insights for realistic materials. 
In this direction, there is a recent work \cite{Regnault2022} where a catalogue of the naturally occurring two and three-dimensional stoichiometric materials with flat bands around the Fermi level was provided. A list of 2,379
high-quality flat-band materials was reported, with 345 candidates that possess topological flat bands, a good number of them layered materials. This
is a good starting point to search for more quantum materials with HOVHS and correlated ground states. In addition, materials where Fermi surface pockets can be identified which can be tuned to merge through external stimuli constitute interesting candidates. 

Another research direction where HOVHS may have a crucial role is in the physics of heavy-fermion compounds. There it has been already noted that a number of unusual properties can be the result of Fermi surface topological transitions \cite{Shankar2023}. As an extension, the unusual exponents that characterize certain physical quantities could result from the formation of HOVHS which needs to be systematically checked.

As we explained above, one reason for the interest in nearly flat bands is that in their presence 
there is the possibility of fascinating new ground states due to interaction effects, e.g., 
ranging from fractional Chern insulators \cite{Neupert_PRL_2011, Sun_Gu_Katsura_DasSarma_PRL_2011, Sheng_NatComms_2011, Regnault_PRX_2011, Parameswaran_etal_2013} to high-temperature superconductivity \cite{Kopnin_Heikkila_Volovik_2011}. 
Multi-band systems offer additional possibilities with qualitatively novel phenomena. 
For example, it was shown that coincident Van Hove points from more than one band facilitate scattering channels that would be forbidden in a single-band case \cite{yu2024altermagnetism}, which can make the system unstable towards altermagnetic or inter-band pairing instabilities. 
The influence of multiple bands also plays an important role for the quantum geometric contribution to the 
superfluid weight \cite{Peotta2015} or excitonic effects \cite{HOvHS3}.  
It will be interesting to expand the investigation of materials with flattened but finite dispersion in these directions. 
Apart from these, there is the possibility to connect HOVHS with D-branes in a more abstract sense, given the connection of HOVHS to Fermi surface topological transitions \cite{Horava_PRL_2005}.

\begin{summary}[SUMMARY POINTS]
\begin{enumerate}
\item HOVHS play important role in correlated quantum materials due to large enhancement of interactions compared to kinetic energy.
\item 
HOVHS can arise from flat bands by different perturbations.
\item HOVHS and Fermi surface topological transitions are closely connected. In many systems pieces of Fermi surfaces merge at high symmetry points of the BZ, providing HOVHS.
\item There are experimental signatures of HOVHS in phase formation, thermodynamic and transport properties. 
\item There is a complete classification scheme in two-dimensions based on catastrophe theory with characteristic exponent for the diverging DOS for each singularity. There is also a developed method to diagnose and analyse HOVHS.
\item Particle-particle and particle-hole susceptibilities becomes equivalently singular at a HOVHS leading to an inevitable competition of orders.
\item There is evidence of HOVHS in kagom\'e metals, moir\'e materials and strontium ruthenates.
\end{enumerate}
\end{summary}

\begin{issues}[FUTURE ISSUES]
\begin{enumerate}
\item What is the role of HOVHS in phase formation under certain conditions (role of orbitals, spin, other parts of the Fermi surface, transport properties, external parameters)? 
\item Is it possible to classify the perturbations that lead to specific HOVHS in flat bands?
\item Can we discover and study new materials with desired properties based on HOVHS?
\end{enumerate}
\end{issues}

\section*{DISCLOSURE STATEMENT}
The authors are not aware of any affiliations, memberships, funding, or financial holdings that
might be perceived as affecting the objectivity of this review. 

\section*{ACKNOWLEDGMENTS}
We gratefully acknowledge discussions with Anirudh Chandrasekaran, Claudio Chamon, Andrey Chubukov, Michael Scherer, and Andreas Schnyder. JJB gratefully acknowledges funding from the UK Engineering and Physical Sciences Research Council (EPSRC) via Grants No EP/T034351/1 and No EP/X012557/1.

\bibliographystyle{ar-style4}
\bibliography{biblio}

\end{document}